\documentclass[12pt]{report}
\usepackage{amssymb}
\usepackage{amsmath}

\usepackage{graphicx}
\begin{document}
\voffset = -2.9cm
\hoffset = -1.3cm
\def\itm{\newline \makebox[8mm]{}}
\def\ls{\makebox[8mm]{}}
\def\fra#1#2{\frac{#1}{#2}}
\def\fr#1#2{#1/#2}
\def\frl#1#2{\mbox{\large $\frac{#1}{\rule[-0mm]{0mm}{3.15mm} #2}$}}
\def\frn#1#2{\mbox{\normalsize $\frac{#1}{\rule[-0mm]{0mm}{3.15mm} #2}$}}
\def\frm#1#2{\mbox{\normalsize $\frac{#1}{\rule[-0mm]{0mm}{2.85mm} #2}$}}
\def\frn#1#2{\mbox{\normalsize $\frac{#1}{\rule[-0mm]{0mm}{3.15mm} #2}$}}
\def\hs#1{\mbox{\hspace{#1}}}
\def\b{\begin{equation}}
\def\e{\end{equation}}
\def\arccot{\mbox{arccot}}
\def\hn{\hspace{0.7mm}}
\vspace*{6mm}
\makebox[\textwidth][c]
{\large \bf{A solution of the Einstein-Maxwell
equations describing }}
\vspace{-4mm} \newline
\makebox[\textwidth][c]{\large \bf{conformally flat
spacetime outside a charged domain wall}}
\vspace{4mm} \newline
\makebox[\textwidth][c]
{\normalsize \O yvind Gr\o n$^{*}$ and Steinar Johannesen$^*$}
\vspace{1mm} \newline
\makebox[\textwidth][c]
{\scriptsize $*$ Oslo University College, Department of Engineering,
P.O.Box 4 St.Olavs Plass, N-0130 Oslo, Norway}
%
\vspace{6mm} \newline
{\bf \small Abstract} {\small We derive and discuss the physical
interpretation of a conformally flat, static solution of the
Einstein-Maxwell equations. It is argued that it describes a conformally
flat, static spacetime outside a charged spherically symmetric domain wall.
The acceleration of gravity is directed away from the wall in spite of the
positive gravitational mass of the electric field outside the wall, as
given by the Tolman-Whittaker expression. The reason for the repulsive
gravitation is the strain of the wall which is calculated using the
Israel formalism for singular surfaces.}
%
%
\vspace{10mm} \newline
{\bf 1. Introduction}
\vspace{3mm} \newline
B.\hn Bertotti [1] and I.\hn Robinson [2] have found a conformally
flat solution of the Einstein-Maxwell equations. The solution can be
represented by the line element \eqref{e_22} below. N.\hn Tariq and
B.\hn O.\hn J.\hn Tupper [3] have proved that this is the only
conformally flat, static solution of the Einstein-Maxwell equations
in which the energy-momentum tensor has no contribution other than
that of a nonnull stationary electromagnetic field. A corresponding
deduction for null electromagnetic fields has been given by
R.\hn G.\hn McLenaghan, N.\hn Tariq and B.\hn O.\hn J.\hn Tupper [4],
and has also been discussed by M.\hn Cahen and J.\hn Leroy [5].
\itm The physical interpretation of the solution has been discussed by
D.\hn Lovelock [6,7] and P.\hn Dolan [8]. Lovelock [6] argues that the
solution describes spacetime outside a static, massless, charged particle.
He also notes that the solution \eqref{e_22} is different from the
Reissner-Nordstr\"{o}m solution with $m = 0$, although both cause
repulsive gravitation. The reason for the repulsive gravitation
and the difference between the solution \eqref{e_22} and the
Reissner-Nordstr\"{o}m solution were, however, not discussed.
\O.\hn Gr\o n [9] has shown that the reason for the Reissner-Nordstr\"{o}m
repulsion is Poincar\'{e} stresses which are always present in a static,
charged object.
\itm It should be noted, however, that Lovelock's assumption of a massless,
charged particle is not admitted. In order to avoid a naked singularity
the particle must have a charge satisfying
$R_{\mbox{\tiny $Q$}} < 2 R_{\mbox{\tiny $S$}}$,
where $R_{\mbox{\tiny $Q$}}$ is the length defined in equation \eqref{e_52}
and $R_{\mbox{\tiny $S$}}$ is the Schwarzschild radius of the particle.
\itm Dolan [7] has argued that the Bertotti-Robinson-Lovelock solution
does not represent spacetime outside a charged particle. He has imbedded
the curved 4-dimensional spacetime in a 6-dimensional pseudo-Euclidean
space, and claims that this construction shows that the 3-space of the
4-dimensional spacetime is not spherically symmetric. In particular,
he points out that there is no physical singularity at $r = 0$, which
it should be if there was a point charge at $r = 0$.
\itm In the present paper we shall construct a solution of the
Einstein-Maxwell equations where the problem at $r = 0$ is avoided,
so that spherical symmetry of the 3-space can be maintained. This is
obtained by excluding $r = 0$ from the conformally flat solution.
We introduce a charged, spherical shell with center at $r = 0$
with flat spacetime inside it. The mechanical properties of this shell
is determined using Israel's relativistic theory [10] for singular layers.
It turns out that this construction will also explain the repulsive
gravity of the conformally flat solution, and the difference
between this solution and the Reissner-Nordstr\"{o}m solution. We will
also show that an extension of the Tolman-Whittaker expression [11]
of gravitational mass is needed in order to obtain a description
which is generally consistent with the acceleration of gravity obtained
from the geodesic equation.
%
%
\vspace{10mm} \newline
{\bf 2. A conformally flat static spacetime with an electric field}
\vspace{3mm} \newline
We shall consider a spherically symmetric, conformally flat and static
spacetime with a radial electric field. Using conformally flat spacetime
(CFS) coordinates $(T,R,\theta,\phi)$ [12 - 14] the line element can be
written
\begin{equation} \label{e_1}
ds^2 = e^{2 {a}} (-dT^2 + dR^2 + R^2 d\Omega^2)
\mbox{ ,}
\end{equation}
where ${a}$ is a function of $R$ alone and
$d\Omega^2 = d\theta^2 + sin^2 \theta \hs{0.5mm} d\phi^2$.
The field equations take the
form
\begin{equation} \label{e_2}
G^{\mbox{\tiny $T$}}_{\mbox{\tiny $T$}}
= e^{-2 {a}} (2 {a}'' + {a}'^2 + \frl{4}{R} {a}')
= - e^{-4 {a}} \frl{G \hs{0.5mm} Q^2}{4 \pi {\epsilon}_0 R^4}
\mbox{ ,}
\end{equation}
\begin{equation} \label{e_3}
G^{\mbox{\tiny $R$}}_{\mbox{\tiny $R$}}
= e^{-2 {a}} (3 {a}'^2 + \frl{4}{R} {a}')
= - e^{-4 {a}} \frl{G \hs{0.5mm} Q^2}{4 \pi {\epsilon}_0 R^4}
\mbox{ ,}
\end{equation}
\begin{equation} \label{e_4}
G^{\theta}_{\theta}= G^{\phi}_{\phi}
= e^{-2 {a}} (2 {a}'' +  {a}'^2 + \frl{2}{R} {a}')
= e^{-4 {a}} \frl{G \hs{0.5mm} Q^2}{4 \pi {\epsilon}_0 R^4}
\mbox{ ,}
\end{equation}
where $G^{\mu}_{\nu}$ are the components of the Einstein tensor,
${\epsilon}_0$ is the permittivity in vacuum, and $Q = Q(R)$ is the
charge inside a spherical surface with radius $R$ [15].
Subtracting equation \eqref{e_4} from equation \eqref{e_2} we
obtain
\begin{equation} \label{e_5}
e^{2 {a}} {a}' = - \frl{G \hs{0.5mm} Q^2}{4 \pi {\epsilon}_0 R^3}
\mbox{ .}
\end{equation}
Inserting this into equation \eqref{e_3} leads to two different solutions,
either
\begin{equation} \label{e_30}
Q(R) = 0
\end{equation}
or
\begin{equation} \label{e_6}
Q(R) = \pm \sqrt{\frl{4 \pi {\epsilon}_0}{G}} \hs{1.0mm} R e^{{a}}
\mbox{ .}
\end{equation}
\itm From equations \eqref{e_5} and \eqref{e_30} we
find that ${a} = {a}_0$ where ${a}_0$ is a constant.
We may adjust the rate of the coordinate clocks and the length of the
measuring rods so that ${a}_0 = 0$. This means that the solution
$Q(R) = 0$ represents the Minkowski spacetime.
\itm Combining equations \eqref{e_5} and \eqref{e_6} leads to
\begin{equation} \label{e_19}
{a}' = - \frl{1}{R}
\mbox{ .}
\end{equation}
Integration gives
\begin{equation} \label{e_7}
e^{a} = \frl{R_{\mbox{\tiny $Q$}}}{R}
\mbox{ ,}
\end{equation}
where $R_{\mbox{\tiny $Q$}}$ is a positive constant of integration. From
equation \eqref{e_6} we now get
\begin{equation} \label{e_8}
Q(R) = \pm \sqrt{\frl{4 \pi {\epsilon}_0}{G}} \hs{1.0mm}
R_{\mbox{\tiny $Q$}}
\mbox{ .}
\end{equation}
Hence the integration constant
\begin{equation} \label{e_52}
R_{\mbox{\tiny $Q$}} = \sqrt{G / (4 \pi {\epsilon}_0 c^4)} \hs{1.0mm} Q
\end{equation}
represents the length corresponding to the charge $Q$ (where we have
included the velocity of light $c$ in this conversion formula between
charge and length).
Equation \eqref{e_8} implies that the charge density vanishes in the
considered region. Thus the assumption of a conformally flat, static
and spherically symmetric spacetime implies that there is no charge
in this region, although there is a radial electric field there.
Hence there must be a charge inside the conformally flat region.
\itm The line element describing spacetime outside the charge
distribution takes the form
\begin{equation} \label{e_22}
ds^2 = \frl{R_{\mbox{\tiny $Q$}}^2}{R^2}
(-dT^2 + dR^2 + R^2 d\Omega^2)
= \frl{R_{\mbox{\tiny $Q$}}^2}{R^2} (-dT^2 + dR^2)
+ R_{\mbox{\tiny $Q$}}^2 d\Omega^2
\mbox{ ,}
\end{equation}
%
Since the metric is static, the coordinate clocks go at the same
rate everywhere, equal to the rate of the standard clocks at
$R = R_{\mbox{\tiny $Q$}}$. We see that the rate of time as shown by
standard clocks at rest is slower for increasing $R$. This indicates
that the field of gravity points in the direction of increasing $R$.
The Kretschmann curvature scalar is constant in this spacetime,
\begin{equation} \label{e_23}
R^{\mu \nu \alpha \beta} R_{\mu \nu \alpha \beta}
= \frl{8}{R_{\mbox{\tiny $Q$}}^4}
\mbox{ .}
\end{equation}
Hence there is no physical singularity in the spacetime represented by
the line element \eqref{e_22}.
\itm We shall assume that the charge $Q$ is distributed uniformly
on a singular spherical shell with a spacetime described by the line
element \eqref{e_22} outside the shell and Minkowski spacetime inside it.
Furthermore, we choose as a coordinate condition that the radial
coordinate is continuous at the shell. From the line element \eqref{e_22}
it then follows that the CFS radius of the shell is
$R = R_{\mbox{\tiny $Q$}}$.
\itm The geometry of the spacetime outside the charged shell is rather
strange. From the line element \eqref{e_22} we see that the area of a
spherical surface is independent of the radius. One may wonder if this
means that there is no usual kind of spherical symmetry as suggested by
Doland [7]. We shall here keep to the interpretation that the line
element \eqref{e_22} represents a spherically symmetric space
with a rather unusual geometry.
\itm The electric field strength is
$E^{\mbox{\tiny $R$}} = F^{\mbox{\tiny $T$} \mbox{\tiny $R$}}$, where
$F^{\mbox{\tiny $T$} \mbox{\tiny $R$}}$ is a contravariant component of
the electromagnetic field tensor. In an orthonormal basis the electric
field outside the spherical charge distribution \eqref{e_8} decreases as
$R^{-2}$,
\begin{equation} \label{e_27}
F^{\hat{\mbox{\tiny $T$}} \hat{\mbox{\tiny $R$}}}
= \frl{Q}{4 \pi {\epsilon}_0 R^2}
\mbox{ ,}
\end{equation}
The coordinate component of the electric field strength in the
CFS coordinate system is
\begin{equation} \label{e_28}
E^{\mbox{\tiny $R$}}
= \frl{F^{\hat{\mbox{\tiny $T$}} \hat{\mbox{\tiny $R$}}}}
{\sqrt{| \hs{0.3mm} g_{\mbox{\tiny $T$} \mbox{\tiny $T$}} \hs{0.5mm}
g_{\mbox{\tiny $R$} \mbox{\tiny $R$}} \hs{0.3mm} |}}
= e^{-2 {a}} F^{\hat{\mbox{\tiny $T$}} \hat{\mbox{\tiny $R$}}}
= \frl{R^2}{R_{\mbox{\tiny $Q$}}^2} \hs{0.5mm}
F^{\hat{\mbox{\tiny $T$}} \hat{\mbox{\tiny $R$}}}
\mbox{ .}
\end{equation}
Using equations \eqref{e_27} and \eqref{e_28} we obtain
\begin{equation} \label{e_29}
E^{\mbox{\tiny $R$}} = \frl{Q}{4 \pi {\epsilon}_0 R_{\mbox{\tiny $Q$}}^2}
\mbox{ .}
\end{equation}
We then have the surprising result that the electric field strength
outside the charge distribution does not decrease with the distance
from the charge in this spherically symmetric space. The
reason is that the flux of the electric field lines through a spherical
surface about the charge is generally independent of the radius $R$ of the
surface. Since the area of the surface is independent of $R$, it follows
that the electric field strength does not decrease with increasing $R$.
This means that the energy momentum tensor of the electric field is
constant, which is the reason why the right hand sides of the field
equations \eqref{e_2} - \eqref{e_4} are constant.
\itm We shall now introduce a new radial coordinate $\hat{r}$ equal to the
physical radial distance and rename the time coordinate so that $t = T$.
From the line element \eqref{e_22} it then follows that
\begin{equation} \label{e_24}
d\hat{r} = \frl{R_{\mbox{\tiny $Q$}}}{R} \hs{0.9mm} dR
\mbox{ .}
\end{equation}
We assume  that there is Minkowski spacetime inside the shell
and choose as a coordinate condition that the radial coordinate is
continuous at the shell. This means that $\hat{r} = R_{\mbox{\tiny $Q$}}$
corresponds to $R = R_{\mbox{\tiny $Q$}}$. Hence the physical radius
of the shell is equal to $R_{\mbox{\tiny $Q$}}$.
Integration of \eqref{e_24} with this condition gives
\begin{equation} \label{e_46}
\hat{r}
= R_{\mbox{\tiny $Q$}} \left( 1 + \ln \frl{R}{R_{\mbox{\tiny $Q$}}}
\right)
\mbox{ .}
\end{equation}
The inverse transformation is
\begin{equation} \label{e_25}
R = R_{\mbox{\tiny $Q$}} e^{(\hat{r} - R_{\mbox{\tiny $Q$}})
/ R_{\mbox{\tiny $Q$}}}
\mbox{ .}
\end{equation}
In terms of the physical radial coordinate the line element takes the form
\begin{equation} \label{e_26}
ds^2 = - e^{-2 (\hat{r} - R_{\mbox{\tiny $Q$}})
/ R_{\mbox{\tiny $Q$}}} dt^2
+ d\hat{r}^2 + R_{\mbox{\tiny $Q$}}^2 d\Omega^2
\mbox{ .}
\end{equation}
We see that the coordinate clocks go at same rate as a standard clock
at the shell where $\hat{r} = R_{\mbox{\tiny $Q$}}$. The line
element \eqref{e_26} reduces to the Minkowski line element at the
shell, showing that the metric is continuous at the shell.
%
%
%
\newpage
{\bf 3. A general expression for the acceleration of gravity}
\vspace{3mm} \newline
We shall calculate an expression for the acceleration of gravity
valid in a static spherically symmetric space, modifying the formula
given in [11].
\itm Consider a free particle instantaneously at rest in a static

gravitational field. It follows a timelike geodesic curve in spacetime,
with equation
\begin{equation} \label{e_9}
\ddot{x}^{\mu} + \Gamma^{\mu}_{\alpha \beta}
\dot{x}^{\alpha} \dot{x}^{\beta} = 0
\mbox{ ,}
\end{equation}
which in the present case reduces to
\begin{equation} \label{e_10}
\ddot{R} = -
\Gamma^{\mbox{\tiny $R$}}_{\mbox{\tiny $T$} \mbox{\tiny $T$}} \dot{T}^2
\mbox{ ,}
\end{equation}
where a dot indicates differentiation with respect to the proper time
of the particle.
With the line element \eqref{e_1} we have
\begin{equation} \label{e_11}
\Gamma^{\mbox{\tiny $R$}}_{\mbox{\tiny $T$} \mbox{\tiny $T$}} = {a}'
\mbox{ .}
\end{equation}
The four velocity identity here leads to
\begin{equation} \label{e_12}
\dot{T} = |g_{\mbox{\tiny $T$} \mbox{\tiny $T$}}|^{-1/2} = e^{-{a}}
\mbox{ ,}
\end{equation}
giving
\begin{equation} \label{e_13}
\ddot{R} = - e^{-2 {a}} {a}'
\mbox{ .}
\end{equation}
At the horizon of a black hole we have that
$g_{\mbox{\tiny $T$} \mbox{\tiny $T$}} = 0$, making
$\ddot{R}$ diverge at the horizon. Hence the acceleration of gravity
at the surface of a black hole, the surface gravity, is defined as
\begin{equation} \label{e_14}
g = |g_{\mbox{\tiny $T$} \mbox{\tiny $T$}}|^{1/2} \ddot{R}
\mbox{ ,}
\end{equation}
which is finite at the surface of a black hole. This gives
\begin{equation} \label{e_15}
g = - e^{-{a}} {a}'
\mbox{ .}
\end{equation}
Note that $g < 0$ means attractive gravity.
\itm For the metric \eqref{e_1} we get for the following combination of
the mixed components of the Einstein tensor
\begin{equation} \label{e_16}
(G^{\mbox{\tiny $T$}}_{\hs{1.0mm} \mbox{\tiny $T$}}
- G^{\mbox{\tiny $R$}}_{\hs{1.0mm} \mbox{\tiny $R$}}
- G^{\theta}_{\hs{1.0mm} \theta} - G^{\phi}_{\hs{1.0mm} \phi})
\hs{0.5mm} R^2 e^{4 {a}}
= - (2 \hs{0.5mm} R^2 e^{2 {a}} {a}')'
= (2 \hs{0.5mm} R^2 e^{3 {a}} a)'
\mbox{ .}
\end{equation}
Integration from $R_{\mbox{\tiny $Q$}}$ to $R$ gives
\begin{equation} \label{e_17}
2 \hs{0.5mm} R^2 e^{3 {a}(R)} g(R)
= - 2 \hs{0.5mm} R_{\mbox{\tiny $Q$}}^2 e^{2 {a}(R_0)} {a}'(R_0)
+ \int_{R_{\mbox{\tiny $Q$}}}^R
(G^{\mbox{\tiny $T$}}_{\hs{1.0mm} \mbox{\tiny $T$}}
- G^{\mbox{\tiny $R$}}_{\hs{1.0mm} \mbox{\tiny $R$}}
- G^{\theta}_{\hs{1.0mm} \theta} - G^{\phi}_{\hs{1.0mm} \phi})
\hs{0.7mm} \tilde{R}^2 e^{4 {a}(\tilde{R})} d\tilde{R}
\mbox{ .}
\end{equation}
Using the field equations, we get
\begin{equation} \label{e_18}
R^2 e^{3 {a}(R)} g(R)
= - R_{\mbox{\tiny $Q$}}^2 e^{2 {a}(R_0)} {a}'(R_0)
+ \frl{\kappa}{2} \int_{R_{\mbox{\tiny $Q$}}}^R
(T^{\mbox{\tiny $T$}}_{\hs{1.0mm} \mbox{\tiny $T$}}
- T^{\mbox{\tiny $R$}}_{\hs{1.0mm} \mbox{\tiny $R$}}
- T^{\theta}_{\hs{1.0mm} \theta} - T^{\phi}_{\hs{1.0mm} \phi})
\hs{0.7mm} \tilde{R}^2 e^{4 {a}(\tilde{R})} d\tilde{R}
\mbox{ .}
\end{equation}
This is essentially a modified expression for the Tolman-Whittaker formula
of active gravitational mass. If the integral is negative, it contributes
with attractive gravity.
%
%
%
\vspace{8mm} \newline
{\bf 4. Repulsive gravitation outside a charged domain wall}
\vspace{3mm} \newline
For the solution given in section 2 the formula \eqref{e_18} reduces to
\begin{equation} \label{e_20}
\frl{R_{\mbox{\tiny $Q$}}}{R} \hs{0.5mm} g
= \frl{1}{R_{\mbox{\tiny $Q$}}}
- \frl{\kappa}{R_{\mbox{\tiny $Q$}}^2} \int_{R_{\mbox{\tiny $Q$}}}^R
\frl{Q(\tilde{R})^2}{\tilde{R}^2} d \tilde{R}
= \frl{1}{R_{\mbox{\tiny $Q$}}} -
\int_{R_{\mbox{\tiny $Q$}}}^R \frl{1}{\tilde{R}^2} d \tilde{R}
\mbox{ ,}
\end{equation}
giving
\begin{equation} \label{e_21}
g = \frl{1}{R_{\mbox{\tiny $Q$}}} > 0
\mbox{ .}
\end{equation}
This means that a free particle accelerates in the positive $R$-direction.
Hence there is repulsive gravitation in spite of the fact that the
contribution from the Tolman-Whittaker expression is attractive.
Note the rather surprising result that the acceleration of gravity
is constant and does not diminish with increasing distance from the
singular shell. This is similar to the behaviour of the electric
field strength which was noted after equation \eqref{e_29}.
\itm In order to investigate the reason for the repulsive gravitation
the Israel formalism for singular shells [10] will be used to find
the mechanical properties of the shell.
\itm The energy momentum tensor of the shell is given by
\begin{equation} \label{e_31}
\kappa S^i_{\hs{0.5mm} j} = [K^i_{\hs{0.5mm} j}]
- {\delta}^i_{\hs{0.5mm} j} [K]
\mbox{ ,}
\end{equation}
where $K^i_{\hs{0.5mm} j}$ are the components of the extrinsic curvature
tensor of the shell, $K = K^i_{\hs{0.5mm} i}$ and $[T] = T_+ - T_-$,
where $T_+$ and $T_-$ are the values of $T$ outside and inside the surface
respectively.
We consider the line element
\begin{equation} \label{e_32}
ds^2 = -e^{\alpha} dt^2 + e^{\beta} dr^2
+ e^{\gamma} (d\theta^2 + sin^2(\theta) d\phi^2)
\mbox{ ,}
\end{equation}
where $\alpha$, $\beta$ and $\gamma$ are functions of $r$.
The unit normal vector to a spherical surface about the origin is
given by
\begin{equation} \label{e_33}
\mbox{\bf n} = e^{-\frac{\beta}{2}} \hs{0.5mm} \mbox{\bf e}_r
\mbox{ .}
\end{equation}
The covariant components of the extrinsic curvature tensor are given by
\begin{equation} \label{e_34}
K_{ij} = - n_{i; \hs{0.3mm} j}
= -n_{i,j} + n_{\lambda} \Gamma^{\lambda}_{\hs{0.9mm} ij}
\mbox{ ,}
\end{equation}
where Latin indices run through the surface coordinates $t$,
$\theta$ and $\phi$, and Greek indices run through the four
spacetime coordinates. The first term $n_{i,j}$ vanishes on the surface,
giving
\begin{equation} \label{e_35}
K_{ij} = - \frl{1}{2} n^r g_{ij,r}
= - \frn{1}{2} e^{-\frac{\beta}{2}} g_{ij,r}
\mbox{ .}
\end{equation}
This gives
\begin{equation} \label{e_36}
K^t_{\hs{0.5mm} t}
= - \frn{1}{2} e^{-\frac{\beta}{2}} \alpha_{,r}
\end{equation}
and
\begin{equation} \label{e_37}
K^{\theta}_{\hs{0.5mm} \theta}
= K^{\phi}_{\hs{0.5mm} \phi}
= - \frn{1}{2} e^{-\frac{\beta}{2}} \gamma_{,r}
\mbox{ .}
\end{equation}
Equations \eqref{e_31}, \eqref{e_36} and \eqref{e_37} give
\begin{equation} \label{e_38}
\kappa S^t_{\hs{0.5mm} t}
= - 2 [K^{\theta}_{\hs{0.5mm} \theta}]
= [e^{-\frac{\beta}{2}} \gamma_{,r}]
\end{equation}
and
\begin{equation} \label{e_39}
\kappa S^{\theta}_{\hs{0.5mm} \theta}
= \kappa S^{\phi}_{\hs{0.5mm} \phi}
= - [K^t_{\hs{0.5mm} t}] - [K^{\theta}_{\hs{0.5mm} \theta}]
= \frl{1}{2} [e^{-\frac{\beta}{2}} (\alpha_{,r} + \gamma_{,r})]
\mbox{ .}
\end{equation}
\itm We shall now apply these formulae to the spherical shell defined
above. Then $\alpha_+ = \beta_+ = 2 {a}$ and $\gamma_+ = 2({a} + \ln R)$
outside the shell, where ${a}$ is given by equation \eqref{e_7},
and $\alpha_- = \beta_- = 0$ and $\gamma_- = R^2$ inside it. This gives
\begin{equation} \label{e_42}
\kappa S^{\mbox{\tiny $T$}}_{\hs{0.9mm} \mbox{\tiny $T$}}
= \kappa S^{\theta}_{\hs{0.9mm} \theta}
= \kappa S^{\phi}_{\hs{0.9mm} \phi}
= - \frl{2}{R_{\mbox{\tiny $Q$}}}
\mbox{ .}
\end{equation}
Surprisingly, the mass density and strain of the shell are independent
of its radius. We see that the components of the energy momentum tensor
of the singular shell may be written
\begin{equation} \label{e_47}
S^i_{\hs{0.9mm} j} = - \sigma {\delta}^i_{\hs{0.9mm} j}
\mbox{ ,}
\end{equation}
which characterizes a domain wall [16]. Hence the energy momentum tensor
in equation \eqref{e_42} describes a charged, spherical domain wall with
mass density
\begin{equation} \label{e_44}
\sigma = \frl{2}{\kappa R_{\mbox{\tiny $Q$}}}
\mbox{ ,}
\end{equation}
showing that the shell has a positive mass
$M = 4 \pi \sigma R_{\mbox{\tiny $Q$}}^2 =
R_{\mbox{\tiny $Q$}} / G$, i.e. the mass of the shell is proportional
to its charge.
This means that it is not possible to change the charge of the shell
without simultaneously changing its mass. In particular, it is not
possible to neutralize it, which is the reason why the line element of
the spacetime outside the shell does not contain independent parameters
for the mass and the charge, as in the Reissner-Nordstr\"{o}m spacetime.
In fact the relationship $R_{\mbox{\tiny $Q$}} = GM$ corresponds to a
maximally charged Reissner-Nordstr\"{o}m black hole.
Equation \eqref{e_47} shows that
\begin{equation} \label{e_48}
S^{\theta}_{\hs{0.9mm} \theta}
= S^{\phi}_{\hs{0.9mm} \phi} = - \sigma
\mbox{ .}
\end{equation}
This means that there is a strain in the shell equal to minus its
mass density. We can now understand the reason for the repulsive
gravitation which was noted after equation \eqref{e_21}. As applied
to the domain wall the Tolman-Whittaker formula gives
\begin{equation} \label{e_58}
g(R_{\mbox{\tiny $Q$}}) =
\frl{\kappa}{2} (S^{\mbox{\tiny $T$}}_{\hs{0.9mm} \mbox{\tiny $T$}}
- S^{\theta}_{\hs{0.9mm} \theta}
- S^{\phi}_{\hs{0.9mm} \phi})
= \frl{1}{R_{\mbox{\tiny $Q$}}}
\end{equation}
in accordance with equation \eqref{e_21}, showing that the domain wall
repels gravitationally the surrounding matter [17]. The energy
of the electric field outside the charged domain wall has positive
gravitational mass which causes attractive gravitation. But the strain
of the domain wall causes repulsive gravitation which dominates over
the attractive gravitation of the electric field at all distances from
the domain wall.
%
%
%
\vspace{10mm} \newline
{\bf 5. Comparison with the Reissner-Nordstr\"{o}m metric}
\vspace{3mm} \newline
The Reissner-Nordstr\"{o}m spacetime is usually represented
by the line element
\begin{equation} \label{e_53}
ds^2 = - \left( 1 - \frl{R_{\mbox{\tiny $S$}}}{r}
+ \frl{R_{\mbox{\tiny $Q$}}^2}{r^2}
\right) dt^2 + \left( 1 - \frl{R_{\mbox{\tiny $S$}}}{r}
+ \frl{R_{\mbox{\tiny $Q$}}^2}{r^2}
\right)^{\hs{-1.0mm} -1} dr^2 + r^2 d\Omega^2
\mbox{ ,}
\end{equation}
where $R_{\mbox{\tiny $S$}} = 2GM$ is the Schwarzschild radius of the
mass $M$ in the Reissner-Nordstr\"{o}m metric. A maximally charged
Reissner-Nordstr\"{o}m spacetime has
$R_{\mbox{\tiny $S$}} = 2 R_{\mbox{\tiny $Q$}}$, giving
\begin{equation} \label{e_54}
ds^2 = - \left( 1 - \frl{R_{\mbox{\tiny $Q$}}}{r} \right)^{\hs{-0.4mm} 2}
dt^2
+ \left( 1 - \frl{R_{\mbox{\tiny $Q$}}}{r} \right)^{\hs{-0.8mm} -2} dr^2
+ r^2 d\Omega^2
\mbox{ .}
\end{equation}
In order to compare the solution \eqref{e_22} with the
maximally charged Reissner-Nordstr\"{o}m spacetime, we shall
express the line element \eqref{e_22} in coordinates $(r,t)$ where
$g_{tt} \hs{0.5mm} g_{rr} = -1$.
From the line element \eqref{e_22} we then have
\begin{equation} \label{e_59}
dr = \frl{R_{\mbox{\tiny $Q$}}^2}{R^2} \hs{0.9mm} dR
\mbox{ .}
\end{equation}
Integration with the condition that $r = R_{\mbox{\tiny $Q$}}$ corresponds
to $R = R_{\mbox{\tiny $Q$}}$ gives
\begin{equation} \label{e_55}
r = R_{\mbox{\tiny $Q$}} \left( 2 - \frl{R_{\mbox{\tiny $Q$}}}{R} \right)
\mbox{ ,}
\end{equation}
mapping the region $R \ge R_{\mbox{\tiny $Q$}}$ onto the bounded
region $R_{\mbox{\tiny $Q$}} \le r < 2 R_{\mbox{\tiny $Q$}}$.
The inverse transformation is
\begin{equation} \label{e_56}
R = R_{\mbox{\tiny $Q$}} \left( 2 - \frl{r}{R_{\mbox{\tiny $Q$}}}
\right)^{\hs{-0.8mm} -1}
\mbox{ .}
\end{equation}
In terms of the radial coordinate $r$ the line element \eqref{e_22}
takes the form
\begin{equation} \label{e_57}
ds^2 = - \left( 2 - \frl{r}{R_{\mbox{\tiny $Q$}}} \right)^{\hs{-0.4mm} 2}
dt^2
+ \left( 2 - \frl{r}{R_{\mbox{\tiny $Q$}}} \right)^{\hs{-0.8mm} -2} dr^2
+ R_{\mbox{\tiny $Q$}}^2 d\Omega^2
\mbox{ .}
\end{equation}
There is a certain similarity between the line elements \eqref{e_57}
and \eqref{e_54}, but it is actually more remarkable that they represent
different spacetimes that cannot be related to each other by a coordinate
transformation. This was already noted by Lovelock [7] in connection with
the Jebsen-Birkhoff theorem [18,19] and the line element \eqref{e_22}.
\itm One may wonder why the spacetime outside the charged spherical
domain wall is different from the Reissner-Nordstr\"{o}m spacetime.
After all, both are spaces outside a charged spherical body. The
difference is in the mechanical properties of the bodies containing
the mass and charge appearing in the metrics.
\itm Applying the Israel formalism to a singular spherical shell
having radius $r = r_0$, with Reissner-Nordstr\"{o}m spacetime
outside and Minkowski spacetime inside, we find that the components
of the energy momentum tensor of the shell are
\begin{equation} \label{e_49}
S^t_{\hs{0.9mm} t}
= - \frl{2}{\kappa r_0} \left[ 1 -
\left( 1 - \frl{R_{\mbox{\tiny $S$}}}{r_0} +
\frl{R_{\mbox{\tiny $Q$}}^2}{r_0^2} \right)^{\hs{-0.8mm} 1/2}
\hs{0.5mm} \right]
\end{equation}
and
\begin{equation} \label{e_43}
S^{\theta}_{\hs{0.9mm} \theta}
= S^{\phi}_{\hs{0.9mm} \phi}
= - \frac{1}{\kappa r_0} \left[ 1 -
\left( 1 - \frl{R_{\mbox{\tiny $S$}}}{2r_0} \right)
\left( 1 - \frl{R_{\mbox{\tiny $S$}}}{r_0}
+ \frl{R_{\mbox{\tiny $Q$}}^2}{r_0^2} \right)^{\hs{-0.8mm} -1/2}
\hs{0.5mm} \right]
\mbox{ .}
\end{equation}
Hence the shell has a positive mass density and is strained. In the
Newtonian limit
\begin{equation} \label{e_50}
S^t_{\hs{0.9mm} t} = - \frl{M}{4 \pi r_0^2}
\mbox{ .}
\end{equation}
Since $S^t_{\hs{0.9mm} t}$ is interpreted as minus the mass density
$\sigma$, we get
\begin{equation} \label{e_51}
M = 4 \pi r_0^2 \sigma
\mbox{ ,}
\end{equation}
showing that the mass of the Reissner-Nordstr\"{o}m metric comes
from the spherical shell.
\itm This shall now be compared with the solution above, where the
singular shell is a spherical domain wall with a charge corresponding
to a maximally charged source of the Reissner-Nordstr\"{o}m spacetime
$R_{\mbox{\tiny $S$}} = 2 R_{\mbox{\tiny $Q$}}$. In the case where this
source has the same radius $r_0 = R_{\mbox{\tiny $Q$}}$ as the domain
wall, the components of its energy momentum reduce to
\begin{equation} \label{e_45}
S^t_{\hs{0.9mm} t} = - \frl{2}{\kappa R_{\mbox{\tiny $Q$}}}
\mbox{\hspace{2mm} , \hspace{3mm}}
S^{\theta}_{\hs{0.9mm} \theta}
= S^{\phi}_{\hs{0.9mm} \phi} = 0
\mbox{ .}
\end{equation}
Hence the spherical shell inside the Reissner-Nordstr\"{o}m spacetime
has the same mass density as the domain wall given in equation \eqref{e_44},
but it has a vanishing strain. This difference in the mechanical properties
of the shells is the reason for the different geometries of the spacetimes
outside the shells.
%
%
%
%
\vspace{5mm} \newline
{\bf 6. Conclusion}
\vspace{3mm} \newline
About fifty years ago Bertotti and Robinson found a solution of the
Einstein-Maxwell equations representing a static, conformally flat
spacetime with a uniform electromagnetic field. There has been no
general agreement as to the physical interpretation of the solution,
which is most simply represented by the line element \eqref{e_22}
above.
\itm It was presented by Robinson as a spherically symmetric solution
of the field equations, while Bertotti writes that the main qualitative
physical feature of the solution is the anisotropy of space. Furthermore,
while the Reissner-Nordstr\"{o}m solution contains two physical parameters,
one representing mass and the other charge, this solution contains only
one parameter which represents the charge producing the electric field.
\itm Lovelock [6] shows from the line element \eqref{e_22} that some of the
components of the Riemann curvature tensor are proportional to $1 / r^2$ and
hence claims that spacetime has a physical singularity at $r = 0$. Because
of this and the fact that the line element contains only one physical
parameter he concludes [7] that the line element \eqref{e_22} corresponds
to the spacetime outside a massless point charge at rest at $r = 0$.
Lovelock also notes that there is repulsive gravity in this spacetime, i.e.
a free neutral particle is accelerated away from the charge at $r = 0$.
This is said to be an unphysical property of the solution, and
Lovelock points out that the solution shares this property with the
Reissner-Nordstr\"{o}m solution with $m = 0$.
\itm We have given a new physical interpretation of the solution which
explains its seemingly unphysical properties. Firstly we have shown
that the Kretschmann curvature scalar is constant, which means that
there is indeed only a coordinate singularity at $r = 0$. Using Israel's
relativistic theory of singular shells we have constructed a physical
system where this coordinate singularity does not appear. It consists of
a charged domain wall with Minkowski spacetime inside the wall and
Robinson-Bertotti-Lovelock spacetime outside it. The charge of the wall
corresponds to that of a maximally charged Reissner-Nordstr\"{o}m
spacetime, which explains that there is only one physical parameter in the
solution. The Tolman-Whittaker expression for the gravitational mass
density of the wall is negative, which means that the repulsive character
of the conformally flat spacetime is due to the strain of the domain wall.
\vspace{5mm} \newline
{\bf References}
%
\begin{enumerate}
\item B.\hn Bertotti, \textit{Uniform Electromagnetic Field in the Theory
of General Relativity}, Phys.\hn Rev. \textbf{116}, 1331 - 1333 (1959).

\item I.\hn Robinson, \textit{A Solution of the Maxwell-Einstein
Equations}, Bull.\hn Acad.\hn Pol.\hn Sci.\hn Ser.\hn Sci.\hn
Math.\hn Astr.\hn Phys. \textbf{7}, 351 - 352 (1959).

\item N.\hn Tariq and B.\hn O.\hn J.\hn Tupper, \textit{The uniqueness of
the Bertotti-Robinson electromagnetic universe}, J.\hn Math.\hn Phys.
\textbf{15}, 2232 - 2235 (1974).

\item R.\hn G.\hn McLenaghan, N.\hn Tariq and B.\hn O.\hn J.\hn Tupper,
\textit{Conformally flat solutions of the Einstein-Maxwell equations for
null electromagnetic fields}, J.\hn Math.\hn Phys. \textbf{16}, 829 - 831
(1975).

\item M.\hn Cahen and J.\hn Leroy, \textit{Exact Solutions of
Maxwell-Einstein Equations}, J.\hn Math.\hn Mech. \textbf{16},
501 - 508 (1966).

\item D.\hn Lovelock, \textit{Weakened Field Equations in General
Relativity Admitting an 'Unphysical' Metric}, Commun.\hn math.\hn Phys.
\textbf{5}, 205 - 214 (1967).

\item D.\hn Lovelock, \textit{A Spherically Symmetric Solution of the
Maxwell-Einstein Equations}, Commun.\hn math.\hn Phys. \textbf{5},
257 - 261 (1967).

\item P.\hn Dolan, \textit{A Singularity Free Solution of the
Maxwell-Einstein Equations}, Commun.\hn math.\hn Phys. \textbf{9},
161 - 168 (1968).

\item \O .\hn Gr\o n, \textit{Poincar\'{e} Stress and the
Reissner-Nordstr\"{o}m Repulsion}, Gen.\hn Rel.\hn Grav. \textbf{20},
123 - 129 (1988).

\item W.\hn Israel, {\it Singular Hyperfaces and Thin Shells in General
Relativity}, Il Nuovo Cimento {\bf 44 B}, 1 - 14 (1966) and
{\bf 48 B}, 1 - 1 (1967).

\item \O .\hn Gr\o n, {\it Repulsive gravitation and electron models},
Phys.\hn Rev. D {\bf 31} 2129 - 2131 (1985).

\item \O .\hn Gr\o n and S.\hn Johannesen, \textit{FRW Universe Models in
Conformally Flat Spacetime Coordinates. I: General Formalism},
Eur.\hn Phys.\hn J.\hn Plus \textbf{126}, 28 (2011).

\item \O .\hn Gr\o n and S.\hn Johannesen, \textit{FRW Universe Models in
Conformally Flat Spacetime Coordinates. II: Universe models with
negative and vanishing spatial curvature},
Eur.\hn Phys.\hn J.\hn Plus \textbf{126}, 29 (2011).

\item \O .\hn Gr\o n and S.\hn Johannesen, \textit{FRW Universe Models in
Conformally Flat Spacetime Coordinates. III: Universe models with
positive spatial curvature},
Eur.\hn Phys.\hn J.\hn Plus \textbf{126}, 30 (2011).

\item A.\hn Melfo and H.\hn Rago, {\it Conformally flat solutions to the
Einstein-Maxwell equations}, Astrophysics and Space Science
{\bf 193}, 9-15 (1992).

\item J.\hn Ipser and P.\hn Sikivie, \textit{Gravitationally repulsive
domain wall}, Phys.\hn Rev.\hn D \textbf{30}, 712 - 719 (1984).

\item A.\hn Barnaveli and M.\hn Gogberashvili, \textit{Gravitational
repulsion of spherical domain walls}, Theoretical and Mathematical
Physics \textbf{100}, 1023 - 1029 (1994).

\item J.\hn T.\hn Jebsen, \textit{\"{U}ber die allgemeinen
kugelsymmetrischen L\"{o}sungen der Einsteinschen Gravitationsgleichungen
im Vakuum}, Arkiv f\"{o}r Matematik, Astronomi och Fysik, 15, Nr. 18 (1921).
Translated to English and reprinted in Gen.\hn Rel.\hn Grav.\hn \textbf{37},
2253 - 2259 (2005).

\item G.\hn D.\hn Birkhoff, \textit{Relativity and Modern Physics},
Harvard Univ.\hn Press, Cambridge, U.\hn S.\hn A. (1923).

\end{enumerate}
\end{document}